\documentclass[referee]{aa}
\usepackage{graphics}

\begin{document}

\title{Nature of Clustering in the
Las Campanas Redshift Survey}

\author{Somnath Bharadwaj\inst{1} 
\and A. K. Gupta\inst{2}
\and T. R. Seshadri\inst{3}}
\institute{Department of Physics and
Meteorology,  and,  Center for Theoretical Studies, \\  
I. I. T. Kharagpur 721 302, INDIA.\\ 
{\em e-mail:} ~somnath@phy.iitkgp.ernet.in
\and J. K. Institute of Applied Physics, Allahabad University, \\
Allahabad 221002, INDIA. \\
{\em e-mail:} ~ashok@mri.ernet.in
\and Mehta Research Institute, Chhatnag Road, Jhusi, \\
Allahabad 211019, INDIA. \\
{\em e-mail:} ~seshadri@mri.ernet.in}
\titlerunning{Nature of Clustering in LCRS}
\maketitle

\begin{abstract}
We have carried out a multi-fractal analysis of the distribution of
galaxies in the three Northern slices of the Las Campanas  Redshift
Survey. Our method takes into account the selection effects and the
complicated geometry of the survey. In this analysis we have studied 
the  scaling properties of the distribution of galaxies on length scales
from  $20 h^{-1} {\rm Mpc}$ to $200 h^{-1} {\rm Mpc}$. Our main
results are: (1) The distribution of  galaxies exhibits  a
multi-fractal scaling behaviour over the  scales   $20 h^{-1} {\rm
Mpc}$  to  $80 h^{-1} {\rm Mpc}$, and, (2) the distribution is homogeneous
on the scales $80h^{-1} {\rm Mpc}$  to  $200 h^{-1} {\rm Mpc}$. We
conclude that the universe is homogeneous at large scales and the
transition to homogeneity occurs somewhere in the range
$80 h^{-1} {\rm Mpc}$  to  $100 h^{-1} {\rm Mpc}$. 
\end{abstract}

\keywords{Galaxies --- Clustering}
\section{Introduction}
\label{intro}
Many surveys have been carried out  to chart the positions  of
galaxies in large regions of the  universe around us, and  many more
surveys which go deeper  into the universe are currently underway or
are planned for the future.   These surveys   give us detailed
information  about the  distribution of matter in  the  universe, and
identifying the salient features that characterize this distribution
has been a very important problem in cosmology.  
The  statistical properties,  the  geometry  and the topology are some of
the features that have been used to characterize the  distribution  of
galaxies, and  a  large variety of tools have been developed and used
for this purpose. 

The correlation functions which characterize the statistical
properties of the distributions have been widely applied to quantify
galaxy  clustering. Of the various correlation functions (2-point, 3-
point, etc...) the galaxy-galaxy two point correlation function
$\xi(r)$ is very well determined on small scales (Peebles 1993 and
references therein) and it has been found to have  the form 
\begin{equation}
\xi(r)=\left( \frac{r}{r_0} \right)^{-\gamma} \hspace{0.5cm}
{\rm with} \hspace{0.5cm} \gamma=1.77 \pm 0.04 \hspace{0.5cm} {\rm and}
\hspace{0.5cm} r_0= 5.4 \pm 1  h^{-1} {\rm Mpc}
\end{equation}

This power-law  form of the two point correlation function suggests
that the universe exhibits a scale invariant behaviour on small
scales $r<r_0$. The two point correlation function becomes steeper at
larger  scales $r > r_0$. It is, however, not very well
determined on very large scales where the observations are consistent
with the correlation function being equal to zero. 
The standard cosmological model and the correlation function analysis
are both  based on the underlying assumption that the universe is
homogeneous on very large scales and  the indication that the
correlation function vanishes at very large scales  is consistent with
this.

Fractal characterization is another way of quantifying the gross
features of the galaxy distribution. Fractals have been invoked to
describe many physical phenomena which exhibit a scale invariant
behaviour and it is very natural to use fractals to describe the 
clustering of galaxies on small scales where the correlation function
analysis  clearly demonstrates a scale invariant behaviour.

Coleman and Pietronero (\cite{fractal}) applied the fractal analysis to galaxy 
distributions and
concluded that it exhibits  a self-similar 
behaviour up to arbitrarily large scales. Their claim that the fractal
behaviour extends out to arbitrarily large scales implies that the
universe is not homogeneous on any scale and hence it is
meaningless to talk about the mean density of the universe.
These conclusions are in  contradiction with the Cosmological
Principle and the entire framework of cosmology, as we understand
today, will have to be revised if these conclusions are true.

 On the other hand, several others (\cite{martjones,borgani}) have applied the
fractal analysis to arrive at conclusions that  are more in keeping
with the standard cosmological model. They conclude that while the
distribution of  galaxies does exhibit self similarity  and scaling
behaviour, the  scaling behaviour is valid only over a range of
length scales and  the galaxy distribution is homogeneous on very
large scales.   Various other observations including the
angular distribution of radio sources and the X-ray background
testify to the universe being homogeneous on large scales
(\cite{smooth}; \cite{scos}). 

Recent analysis of the ESO slice
project (\cite{guzzo}) also indicates  that  the universe is
homogeneous over large scales. The fractal analysis of volume limited
subsamples of the SSRS2 (\cite{yx}) studies the spatial
behaviour of the conditional density at scales up to $40
h^{-1} \rm Mpc$. Their analysis is unable to conclusively determine 
whether the distribution of galaxies is fractal or homogeneous and
it is  consistent with both the scenarios. A similar
analysis carried out  for the APM-Stromlo survey (\cite{xz}) seems to indicate that the 
distribution of galaxies exhibits a
fractal behaviour with a dimension of $D=2.1 \pm 0.1$ on scales up to
$40 h^{-1} \rm Mpc$.  In a more  recent paper (\cite{X}) the fractal analysis  has been applied to volume limited
subsamples of the Las Campanas  Redshift Survey. This uses  the
conditional density to probe scales up to  $200 h^{-1} \rm Mpc$. They
find evidence for a fractal behaviour with dimension $D \simeq 2$ on
scales up to  $20 \hbox{--} 40 h^{-1} \rm Mpc$. They also conclude that there is
a tendency to homogenization on  larger scales  
($50 \hbox{--} 100 h^{-1} \rm Mpc$)  
where the fractal dimension  has a value  $D \simeq 3$, but the
scatter in the results is too large to conclusively establish
homogeneity and rule out a fractal universe on large scales.

In this paper we study the scaling properties of the galaxy
distribution in the Las Campanas Redshift Survey  (LCRS) (\cite{lcrs}). 
This  is the deepest redshift survey available at
present. Here  we apply  the multi-fractal analysis
(\cite{martjones,borgani}) which is based on a generalization of the
concept of a mono-fractal.  
In a mono-fractal the scaling behaviour of the point distribution 
is the same around each point and the whole distribution is
characterized by a single  scaling index which corresponds to the fractal
dimension. A  multi-fractal allows for a sequence of scaling indices
known as the multi-fractal spectrum  of generalized dimensions. This
allows for the possibility that the scaling behaviour is not the same
around each point. The spectrum of generalized dimensions tells how
the scaling properties of the galaxy distribution changes from the
very dense regions (clusters) to the sparsely populated  regions (voids) 
in the survey.

 In this paper we  compute the spectrum of generalized dimensions
($D_q$ {\em vs} $q$) by  
calculating the Minkowski-Bouligand dimension (\cite{borgani}) for
both volume limited and magnitude limited subsamples of the LCRS.
We also investigate how the spectrum of generalized dimensions depends
on the length scales over which it is measured and whether the
distribution of galaxies in the LCRS exhibits homogeneity on very
large scales or if the fractal nature extends to arbitrarily large
scales. .

We next present a brief outline of the organization of the paper.
Section ~\ref{gendim} describes the method 
we adopt to compute the spectrum of generalized dimensions.
In section ~\ref{survey} we describe the basic features of the LCRS
and discuss the
issues related to the processing of the data so as to bring it into a form
usable for our purpose. 
Section ~\ref{analysis} gives the details of the 
method of analysis specifically in the context of LCRS.
The discussion of the results are presented in  section ~\ref{results}
and the conclusions in section ~\ref{conc}..

In several parts of the analysis it is required to use definite
values for the Hubble parameter  $H_0 (= 100 h {\rm km/s/Mpc})$ and the
decceleration parameter $q_0$, and we have used $h=1$ and $q_0=.5$.

\section{Generalized Dimension}
\label{gendim}
A fractal point distribution is  usually characterized by its
dimension and there exists a large variety of ways in which the
dimension can be defined and measured. Of these possibilities two
which are particularly simple and can be easily applied to a finite
distribution of points are the box-counting dimension and the
correlation dimension. In this section we discuss the ``working
definitions'' of these two quantities that we have adopted  for
analyzing a distribution of a finite number of points. For  more formal
definitions of these dimensions the reader is referred to Borgani
(1995) and references therein. 
The formal definitions usually involve the limit where the number of
particles tends to infinity and they cannot be directly applied to
galaxy distributions. 

We first consider the box-counting dimension.
In calculating the box-counting dimension for a distribution of points,
the space is divided into identical boxes and we count the
number of  boxes which contain at least one point inside them.  We then
progressively reduce the size of the boxes while   counting the number
of boxes with at least one point inside them  at every stage of
this process. This  gives the number of non-empty boxes $N(r)$
as a function of the size of one edge of the  box $r$ at every stage
of the procedure.  If the number of non-empty boxes exhibits  a power-law
scaling as a function of the size of the box i.e.
\begin{equation}
 N(r) \propto r^{D} \label{eq:gd1}
\end{equation}  
we then define $D$  to be the box-counting dimension. 
In practice the  nature of the scaling may be different on
different length  scales and    we  look for a sufficiently  large
range of $r$ over which  $ N(r)$ exhibits a particular scaling
behaviour  and we then use equation  (\ref{eq:gd1}) to obtain the
box-counting dimension valid over those scales. So finally we may get
more than one value of box-counting dimension for the distribution,
each value 
of the box counting dimension being valid over a limited range of
length scales.

To compute the correlation dimension for a point distribution with N
points  we proceed by first  labeling  the points using an index j
which runs from $1$ to $N$.  We then randomly select $M$ of the $N$
points and the index $i$ is used to refer to these $M$ randomly chosen 
points.

For every point $i$, we count the total number of points  which are
within a distance $r$ from the $i^{th}$ point and  this quantity
$n_i(r)$ can be written as 
\begin{equation}
n_i(r) =\sum_{j=1}^{N} {\Theta}(r-\mid {\vec {x_i}}-{\vec {x_j } }\mid)
\label{eq:gd2}
\end{equation}
where $\vec {x_i }$ is the position vector of the $i^{th}$ point
and $\Theta$ is the Heavy-side function. $\Theta=0$ for
$x<0$ and $\Theta=1$ for $x{\ge}0$. We next divide $n_i(r)$ by the
total number of points $N$ to calculate $p_i(r)$, the probability of
finding a point within a distance $r$ from the $i {\rm th}$ point.
We then  average the quantity, $p_i(r)$, over all the $M$  randomly
selected centers to determine the probability of finding a point
within a distance $r$ of another point and  we denote this by
$C_2(r)$  which  is given by,   
\begin{eqnarray}
C_2(r)  =  \frac{1}{M N} \sum_{i=1}^{M} n_i (r) \,. \label{eq:gd3}
\end{eqnarray}

 If the probability $C_2$ exhibits a scaling relation 
\begin{equation}
 C_{2}(r) \propto r^{D_2} \label{eq:gd4}
\end{equation}  
we then define $D_2$  to be the correlation dimension. 

As with the box-counting dimension, the nature of the scaling
behaviour may be different on different length scales and we may then
get more than one value for the correlation dimension, each different
value being valid over a range of scales. 
 
It is very clear that $C_2(r)$ - which is the probability of finding a
point within a sphere of radius $r$ centered on  another point,  is closely
related to the  volume integral of the two point correlation function. In
a situation  where the two point correlation function exhibits a
power-law behaviour $\xi(r)=(r/r_o)^{-\gamma}$ on scales $r<r_0$,  we
expect the correlation dimension to have a value $D_2=3-\gamma$ over
these scales.  

For a mono-fractal the box-counting dimension and the correlation
dimension will be the same, and for a homogeneous, space filling point
distribution they are both equal to the dimension of the ambient
space in which the points are embedded.

The  box-counting dimension and the correlation dimension  quantify
different aspects of the scaling behaviour of a point distribution
and they will have different values in a generic situation. 
The  concept of a
generalized dimension connects these two definitions
and provides a continuous spectrum of dimensions  $D_q$ for  a range of
the parameter $q$. The definition of the Minkowski-Bouligand
dimension $D_q$ (\cite{falconer,feder})
closely follows the definition of the correlation
dimension the only difference being that we use the  $(q-1) {\rm th}$
moment of the galaxy distribution $n_i(r) $ (eq. \ref{eq:gd2})  around
any point. Equation (\ref{eq:gd3}) can then be generalized to  define 
\begin{eqnarray}
C_{q}(r)= {\frac{1}{N M}}\sum_{i=1}^{M}[n_i(<r)]^{q-1}  \label{eq:gd5} \,.
\end{eqnarray}
which is used to define the generalized dimension
\begin{equation}
D_q=\frac{1}{q-1}  \frac {d{\ln}C_{q}(r)}{d{\ln}r} \,. \label{eq:gd6}
\end{equation}

The quantity  $C_{q}(r)$ may exhibit different scaling behaviour over
different ranges of length scales and we will then get more than one
spectrum of generalized dimensions each being valid over a different
range of length scales.  

From equations (\ref{eq:gd5}) and (\ref{eq:gd6}) it is clear that the 
the generalized dimension $D_q$ corresponds to the correlation
dimension at $q=2$. In addition $D_q$ corresponds to the box-counting
dimension at $q=1$. 

For a mono-fractal the generalized dimension is a constant i.e. $D_q=D$
which reflects the fact that for a mono-fractal the point distribution
is characterized by a unique scaling behaviour. For a generic
multi-fractal the values of $D_q$ will be different for different
values of $q$. The positive values of $q$ give more weight-age to the
over-dense regions. Thus, for $q > 0$, $D_q$ probes the scaling
behaviour of the distribution of points in the over-dense regions like
inside clusters etc.  The negative values of $q$, on the other hand,  
give more weight-age to the
under-dense regions and, hence, for negative $q$,  $D_q$ probes the scaling
behaviour of the distribution of points in the under-dense regions like
voids.  

Finally it should be pointed out that the   Minkowski-Bouligand
generalized dimension  $D_q$  is one of the
many possible definitions of a  generalized dimension.
The  minimal spanning tree  used by van der Weygaert and Jones
(\cite{Weygaert}) is another possible method which can be used. 
The   Minkowski-Bouligand generalized dimension has the advantage of
being easy to compute. In addition  the various selection effects
which have to be taken into account when analyzing redshift surveys can
be easily accounted for when determining the Minkowski-Bouligand
generalized dimension  and hence we have chosen this
particular method  for the  multi-fractal characterization of the
galaxy distribution in  LCRS,

\section{A Brief Description of the Survey and the Data.}
\label{survey}
The LCRS consists of 6 alternating slices each subtending
$80^{\circ}$ in right-ascension and  $1.5^{\circ}$ in declination,
3 each in the Northern and Southern Galactic Caps   centered at
$\delta = -3^{\circ}, -6^{\circ}, -12^{\circ}$ and 
$\delta = -39^{\circ}, -42^{\circ}, -45^{\circ}$ respectively. 
The survey extends to a redshift of $\sim .2$ corresponding to $600
h^{-1}{\rm Mpc}$ in the radial direction. The survey contains about  
24000 galaxies distributed with a mean redshift of $z=.1$
corresponding to $300 h^{-1} {\rm Mpc}$.

We next elaborate a little on the shape of the individual
slices. Consider  two cones both with the same axis and with their
vertices at the same point. Let the angle between the first cone and
the axis be   $90^{\circ}-(\delta - .75^{\circ})$ and the second cone
and the axis be  $90^{\circ}-(\delta + .75^{\circ})$ so that the
angle between the two cones is $1.5^{\circ}$.  Next truncate both the
cones at a radial distance of $600 h^{-1} {\rm Mpc}$ from the
vertex. Finally, a slice centered at a declination $\delta$ corresponds
to a $80^{\circ}$ wedge of the region between these two cones. 
The effect of the extrinsic curvature of the cones is small
for the three northern slices and  we have restricted our analysis to
only these three slices for which we have neglected the effect of the
curvature.  

Each slice in the LCRS is made up of $1.5^\circ$ x $1.5^\circ$ fields
some of which were observed with a $50$ object fibre system and others
with a $112$ object fibre system. Of the three northern slices the one
at $\delta=-12^\circ$ is exclusively made up of 112 fibre fields while
the slice at $\delta=-6^\circ$  is mostly 50 fibre, and the slice
at $\delta=-3^\circ$ has got both 50 and 112 fibre fields. 

For each field, redshifts were determined for those galaxies which
satisfy the magnitude limits and the central brightness limits of the
survey. These limits are different for the 50 fibre and the 112 fibre
fields. In addition, for those fields where the number of galaxies
satisfying the  criteria for inclusion in the survey exceeded the
number of fibres, the redshifts were determined for  only a fraction
of the galaxies  in the field. This effect is quantified by the
``galaxy sampling function'' $f$ which varies from field to field
and  is around $80 \%$ for the 112 fibre
fields and around half this number for the 50 fibre fields. In
addition to the field to field variation of the galaxy sampling
function there are two other effects which have to be accounted for when
analyzing the galaxy distribution. They are, 
(1). Apparent Magnitude and Surface
Brightness Incompleteness, and, (2). Central Surface Brightness Selection.
These are quantified by two factors $F$ and $G$, respectively, which
are discussed in detail in Lin {\it et al.} (1996). The survey data
files provide the product of these three factors $sf=f \cdot F \cdot
G$ for each galaxy and the contribution from the $i \, th$
galaxy has to be weighted with the factor  
\begin{equation}
W_i=\frac{1}{f_i \cdot F_i \cdot G_i} \label{eq:s1}
\end{equation}
when analyzing the survey.

The factor $W_i$ discussed above takes into account  the effects
of the field-to-field sampling fraction and the incompleteness as a
function of the apparent magnitude and central surface brightness. 
In addition,  the selection function $s(r)$  has also to be taken into
account, and this depends on both the differential luminosity
function $\phi(M)$ and  the magnitude limits of the survey. The luminosity
function of LCRS has been studied by Lin {\it et al.} (1996) who have
determined  the luminosity function for different sub-samples of LCRS.

They find that the  Schechter form  with the parameters
$M^{\star}=-20.29 + 5 \log(h)$, $\alpha=-0.70$ and   $\phi^{\star}=
0.019  h^{3} {\rm Mpc}^{-3}$ provides a good fit for  the luminosity
function in the absolute magnitude range $-23.0 \ge M  \ge -17.5$.
They have obtained  these parameters from the   analysis of the  combined
Northern and Southern 112 fibre fields  and we shall refer to the
Schechter luminosity function with these set of parameters as the NS112
luminosity function. The analysis of Lin {\it et al.} (1996) shows that
this luminosity function can be used for the Northern 50 fibre fields
in addition to the Northern and Southern 112 fibre fields,  and we have
used the NS112 luminosity function for most of our analysis. 

Lin {\it et al.} (1996) have also separately provided the luminosity function  
determined using just the Northern 112 fibre fields. This has the
Schechter form with the parameters   $M^{\star}=-20.28 + 5 \log(h)$,
$\alpha=-0.75$ and   $\phi^{\star}= 0.018  h^{3} {\rm Mpc}^{-3}$ and
we refer to this as the N112 luminosity function. We have used this
in some of our analysis of the $\delta=-12^{\circ}$ slice  
which contains  only 112 fibre fields. 

The selection function $s(z)$ quantifies the fact that  the fraction of the
galaxies  which are expected to  be included in the survey
varies with the distance from the observer.   For a magnitude limited
survey the  apparent magnitude limits $m_1$ and $m_2$ can be converted
to absolute  magnitude limits $M_1(z)$ and $M_2(z)$ at some redshift
$z$. In  addition if we impose further  absolute magnitude criteria
$M_1 \ge M 
\ge M_2$, then the selection function can be expressed as 
\begin{equation}
s(z)=\int^{min[M_2(z),M_2]}_{max[M_1(z),M_1]} \phi(M) d M 
{\Big /} \int^{M_2}_{M_1} \phi(M) d M \,. \label{eq:s2}
\end{equation}

The apparent magnitude limits are different for the 50 and 112 fibre
fields and we have used the appropriate magnitude limits and the N112/
NS112 luminosity functions  to calculate the selection function 
at the redshift of each of the galaxies. This is then used to
calculate a weight factor for each of the galaxies, and the
contribution of  the $i \, th$ galaxy in the survey  has to be
weighed   by 
\begin{equation}
w_i=\frac{W_i}{ s(z_i)} \,. \label{eq:s3}
\end{equation} 

Another effect that we  have to correct for  arises because of the   
fact that we would like to treat the distribution of galaxies in each
slice as a two 
dimensional distribution. Each slice  consists of galaxies
that are contained within a  thin  conical shell of thickness $1.5^o$
and we construct a two dimensional distribution by collapsing the
thickness of the slice. The thickness of each slice increases 
with the distance from the observer and in order to compensate
for this effect we weigh each galaxy by the inverse of the thickness of
the slice at its red-shift. Taking this effect into account the weight
factor gets modified to    
\begin{equation}
w_i=\frac{W_i}{z_i s(z_i)} \,. \label{eq:s4}
\end{equation} 
which we use to weigh the contribution from the $i \, th $ galaxy  in
the LCRS.

We should also point out that through  the process of flattening the 
conical slices and collapsing its thickness,  the three dimensional
galaxy distribution 
has been converted to a 2-dimensional distribution and the whole of
our multi-fractal analysis  is for a planar 2-dimensional point
distribution.  

In our analysis we have considered various subsamples of  LCRS all
chosen from the 3 Northern slices. In addition to the apparent
magnitude limits of the survey we have imposed further absolute
magnitude and redshift cutoffs to construct both volume and apparent
magnitude limited subsamples whose details are presented in Table I. 

\section{Method of Analysis}
\label{analysis}
We first extract  various subsamples of LCRS using the criteria given
in Table I for each of the subsample. For each subsample  we next
calculate the weight function $w_i$  (equation \ref{eq:s4}) for all
the galaxies in the subsample.  In addition the 3-dimensional
distribution of galaxies in the sub-sample is converted  into a
corresponding  2-dimensional distribution using the steps outlined in
the  previous section and we finally have a collection of $N$ galaxies
distributed over a region of a plane.  

We next  choose $M$ of these galaxies at random and count the   number
of galaxies inside a circle of radius $r$ drawn around each of these
$M$ randomly chosen
galaxies. In determining this we use a modified version of equation 
(\ref{eq:gd2}) where each galaxy in the circle has an extra weight
factor $w_j$ as calculated in the previous section, i.e.  
\begin{equation}
n_i(r) =\sum_{j=1}^{N} w_j {\Theta}(r-\mid {\vec {x_i}}-{\vec {x_j } }\mid)
\label{eq:az1} \,.
\end{equation}

The different moments of this quantity are averaged over the $M$
galaxies to obtain $C_q(r)$ defined in equation (\ref{eq:gd5}) for a
range of $q$. The exercise is repeated with circles of different radii
(different values of $r$) to finally obtain $C_q(r)$ for a large range
of $r$. 

It should be noted that the region  from which the $M$ points can
be chosen at random depends on the size of the circle which we are
considering. For very large values of $r$ a large region around
the boundaries of the survey has to be excluded because  a 
circle of radius $r$ drawn around a galaxy in that region will extend
beyond the boundaries of the survey. As a consequence for large values
of $r$ we do not have many galaxies which can serve as centers, while
for small values of $r$ there are many galaxies which can serve as
centers for circles of radius $r$. For $r$ between $80 h^{-1} {\rm
Mpc}$ to $200 h^{-1} {\rm Mpc}$ we use $M=60$ which is of the same
order as the  the total number of galaxies  available for use as
centers. To estimate the statistical significance of our results at
this range of length-scales we have randomly
divided the 60 centers into independent groups of 20 centers and
repeated the analysis for each of these. We have used the variation in
the results from the different subsamples to estimate the statistical
errors for our results on large scales.   
  In the range $r<80 h^{-1} {\rm Mpc}$ we have used $M=100$
which is only a small fraction of the total number of galaxies
which can possibly serve as centers which is around 1500. At this
range of length-scales it is possible to choose many independent sets
of 100 centers. We have performed the analysis for a large 
number of such sets of 100 centers and  these have been used to estimate
the mean generalized dimension $D_q$ and the statistical errors in the
estimated $D_q$ at small scales. For both the range of length-scales
considered we  have tried the analysis making changes in the number of
centers and  we find that the results do not vary drastically as we
vary the number  of centers used in the analysis.

\begin{figure}[cr1]
\resizebox{7cm}{7cm}{\includegraphics{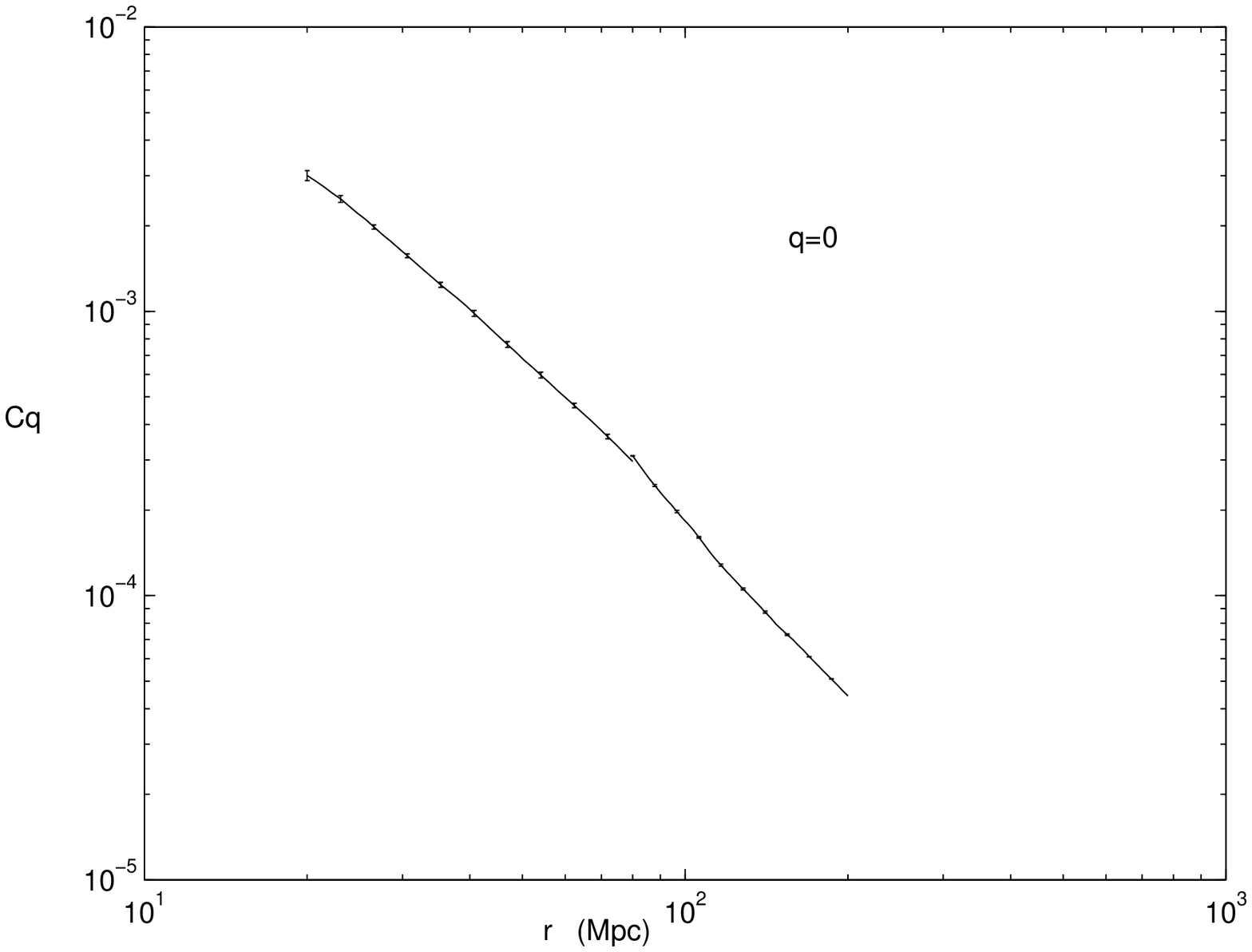}}
\hfill
\caption{This shows $C_q$ (defined in equation \ref{eq:gd5})
as a function of $r$ for  $q=0$  for the subsample d-12.1.}
\label{scaleq0}
\end{figure}

\begin{figure}[cr2]
\resizebox{7cm}{7cm}{\includegraphics{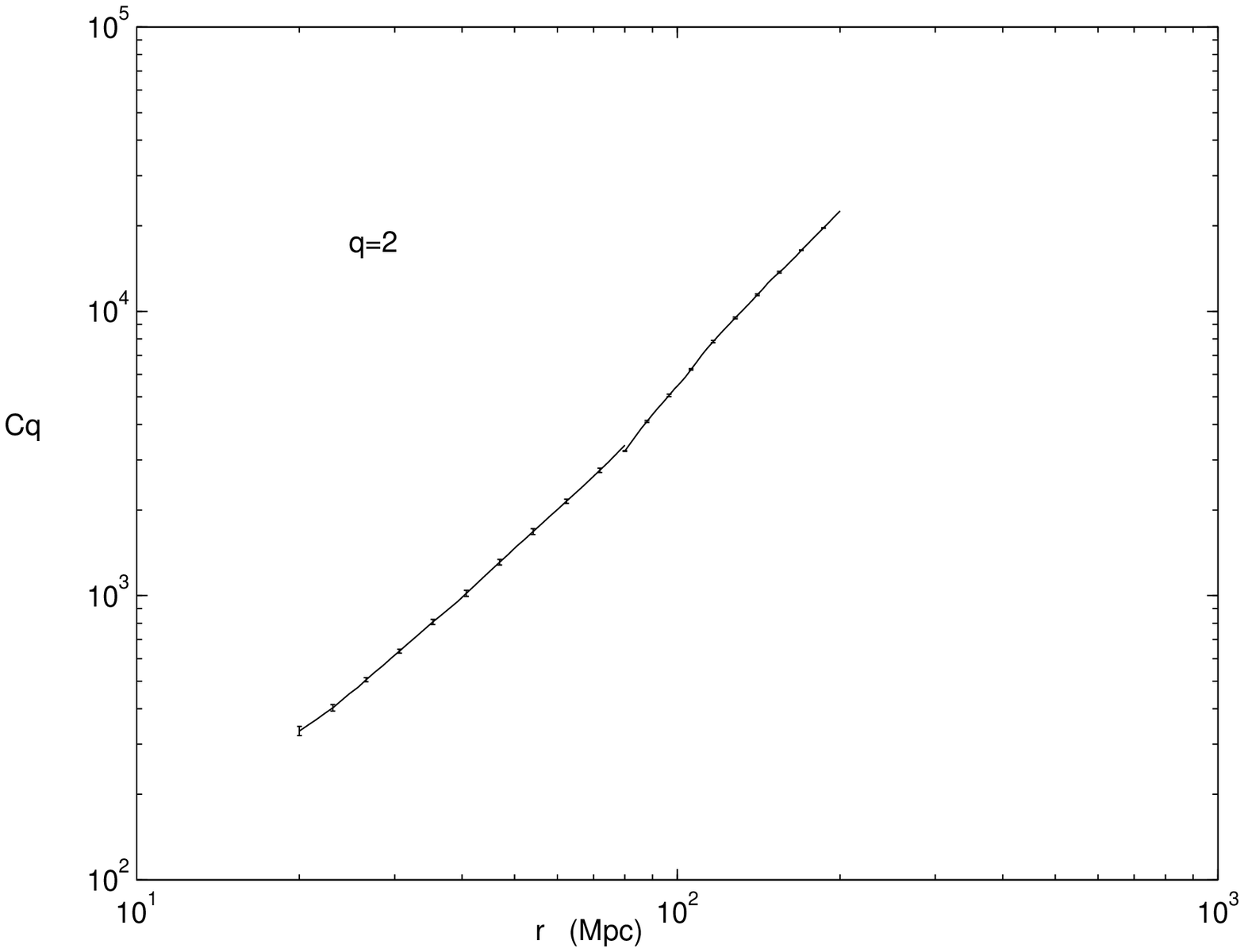}}
\hfill
\caption{This shows $C_q$ (defined in equation \ref{eq:gd5})
as a function of $r$ for  $q=2$  for the subsample d-12.1.}
\label{scaleq2}
\end{figure}

The value of the generalized dimension $D_q$ is determined for a fixed
value of $q$ by looking   at the scaling behaviour of $C_q(r)$ as a
function of $r$ ({\it e.g.} Figures \ref{scaleq0} and \ref{scaleq2}) 
We  have considered $q$ in the range $-10 \le q \le +10$. In principle
we could have considered arbitrarily large (or small) values of $q$
also, but the fact that there are only a finite number of galaxies in
the survey implies that only a finite number of the moments can have
independent information. This point has been discussed in more detail 
by Bouchet {\it et.al.} (\cite{bouchet}). 

In addition to the subsamples of galaxies listed in table
1, we have also carried out our analysis for  mock versions of these
subsamples of galaxies. The mock versions of each subsample contains
the same number of galaxies as the actual subsample. The galaxies in
the mock versions are
selected from a homogeneous random distribution using the same
selection function and geometry as the actual subsample. 
We have carried out the whole analysis for many different random
realizations  of each of the  subsamples listed in Table I. The main
aim of this exercise was to test the reliability of the method of
analysis adopted here. 

\section{Results and Discussion}
\label{results}
We first discuss our analysis of the mock subsamples. Since the
effect of the selection function and the geometry of the slices have
both been included in generating these  subsamples, our analysis of
these subsamples allows us to check how well  these  effects are
being corrected for. In the ideal situation for all the mock
subsamples we should recover a flat spectrum of generalized dimensions
with $D_q=2$ corresponding to a homogeneous point distribution.  
The actual results of the multi-fractal analysis of the mock subsamples
are presented below where we separately discuss the behaviour of $D_q$ at 
small scales  $(r<80 h^{-1} {\rm Mpc})$  and at large scales $(r>80 h^{-1}
{\rm Mpc})$. 

\begin{figure}[mo1]
\resizebox{7cm}{7cm}{\includegraphics{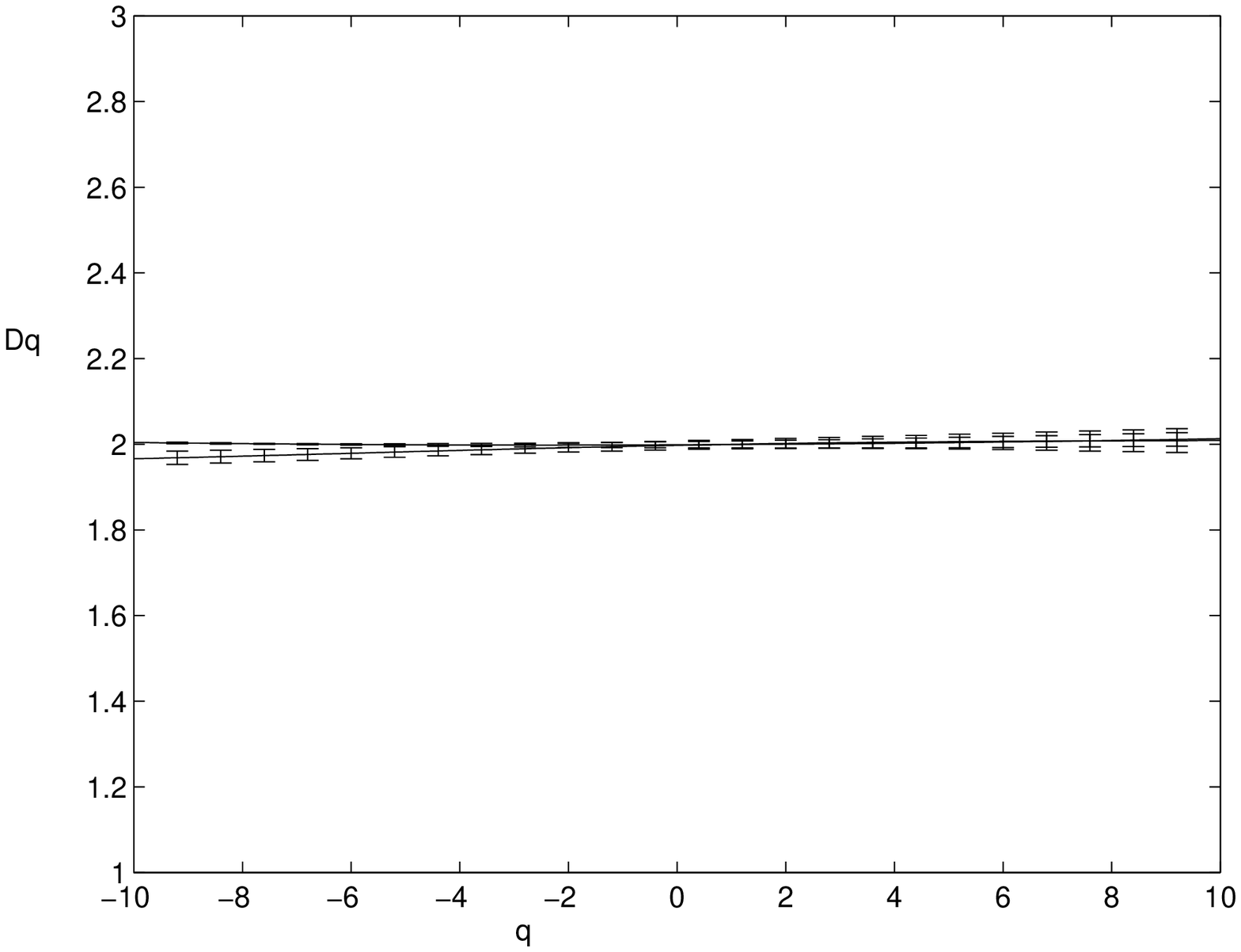}}
\hfill
\caption{The spectrum of generalized dimensions  for  mock subsamples
of  d-12.1 for both small as well as large scales. The curve 
with higher values of $D_q$ at $q=-10$ corresponds to small
scales. The error bars show $1 \hbox{--} \sigma$ statistical errors.}
\label{mock12}
\end{figure}

The results  for mock versions of the subsample d-12.1 are  shown in
figure (\ref{mock12}).   This is a magnitude limited subsample from a slice
that has  only  112 fibre fields  and it contains the largest
number of galaxies.  
We get  a nearly flat spectrum with $D_q=2$ corresponding to a homogeneous
point distribution at both small and large scales. 
Similar results are
also obtained for mock versions of the other subsamples of the
$\delta=-12^\circ$ slice. 

\begin{figure}[mo2]
\resizebox{7cm}{7cm}{\includegraphics{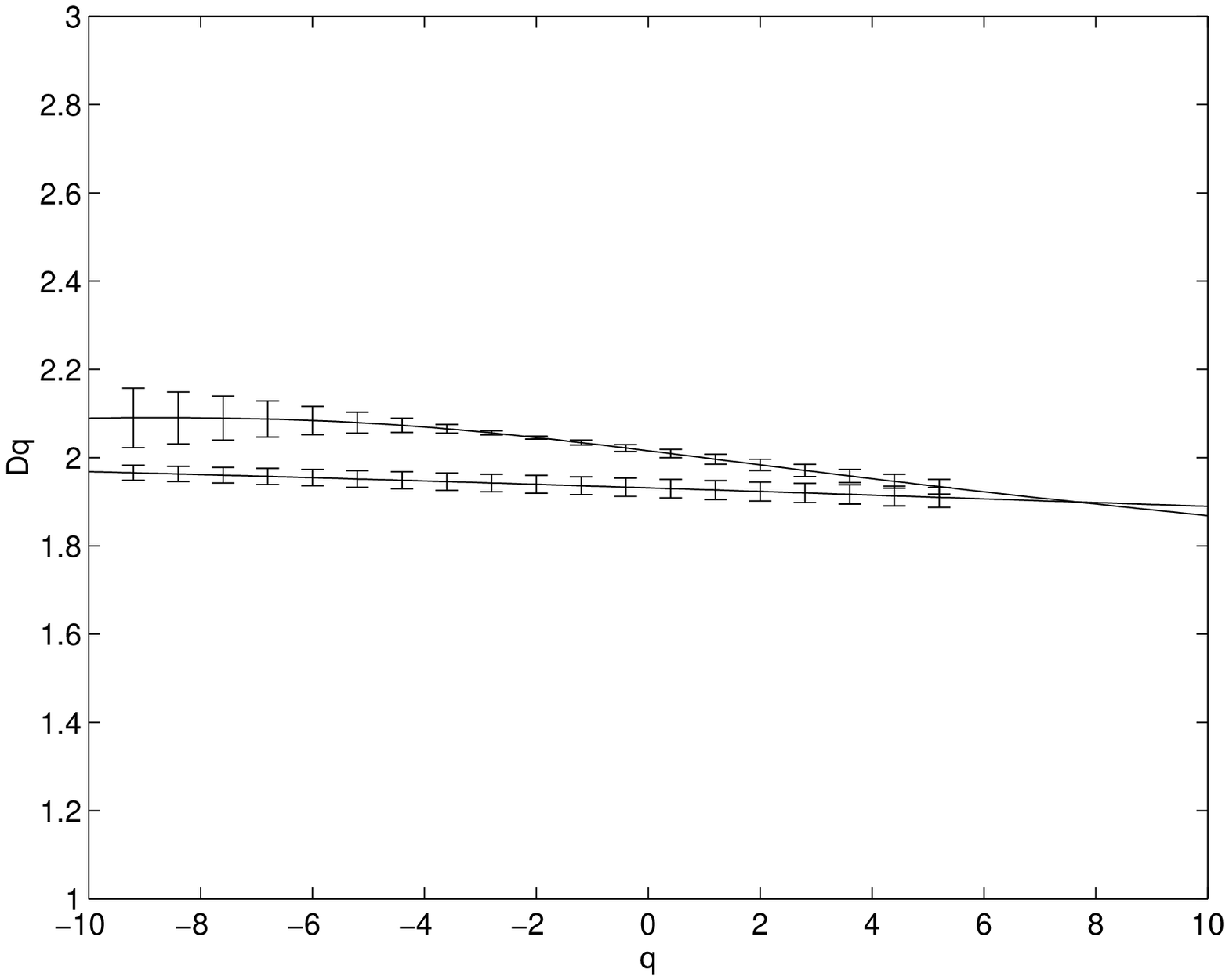}}
\hfill
\caption{The spectrum of generalized dimensions  for  mock subsamples
of  d-03.1 for both small as well as large scales. The curve 
with higher values of $D_q$ at $q=-10$ corresponds to small
scales. The error bars show $1 \hbox{--} \sigma$ statistical errors.}
\label{mock3}
\end{figure}

\begin{figure}[mo3]
\resizebox{7cm}{7cm}{\includegraphics{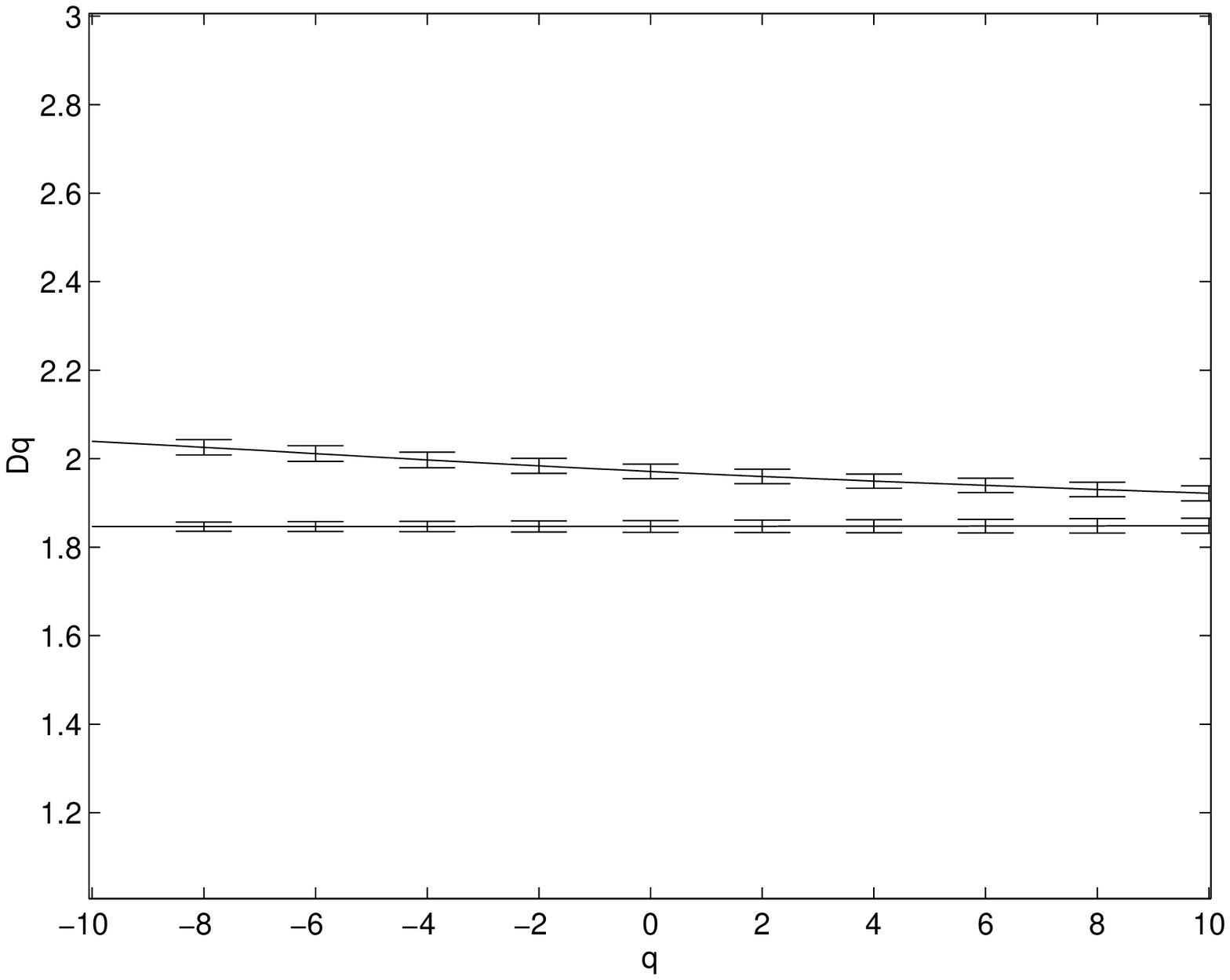}}
\hfill
\caption{The spectrum of generalized dimensions  for  mock subsamples
of  d-06.1 for both small as well as large scales. The curve 
with higher values of $D_q$ at $q=-10$ corresponds to small
scales. The error bars show $1 \hbox{--} \sigma$ statistical errors.}
\label{mock6}
\end{figure}

The analysis of  mock versions of the
subsample d-03.1 which contains both 112 and 50 fibre fields  gives
a spectrum with a weak $q$ dependence (figure ~\ref{mock3}). This
effect is more noticeable at small scales than at large scales.  
The analysis of  mock versions of the d-06.1 subsample (figure
~\ref{mock6}) gives similar results at small scales. At large scales
we get a nearly flat curve with $D_q \simeq 1.8$ This subsample d-06.1
has   mostly 50 fibre fields and it has around  half the number of
galaxies as the  d-12.1 subsample. 

We thus find that the  analysis is most effective for the subsample
from the $\delta=-12^\circ$ slice where $D_q$ shows  very little $q$
dependence and $2.1 \le D_q \le 1.9$. For the other  two  slices we
find a weak $q$ dependence with  $2.2 \le D_q \le  1.8$. 
This clearly demonstrates that our method of multi-fractal analysis
correctly takes into account the different  selection effects and the
complicated sampling and geometry  for all the subsamples that 
we have considered. 

\begin{figure}[ac1]
\resizebox{7cm}{7cm}{\includegraphics{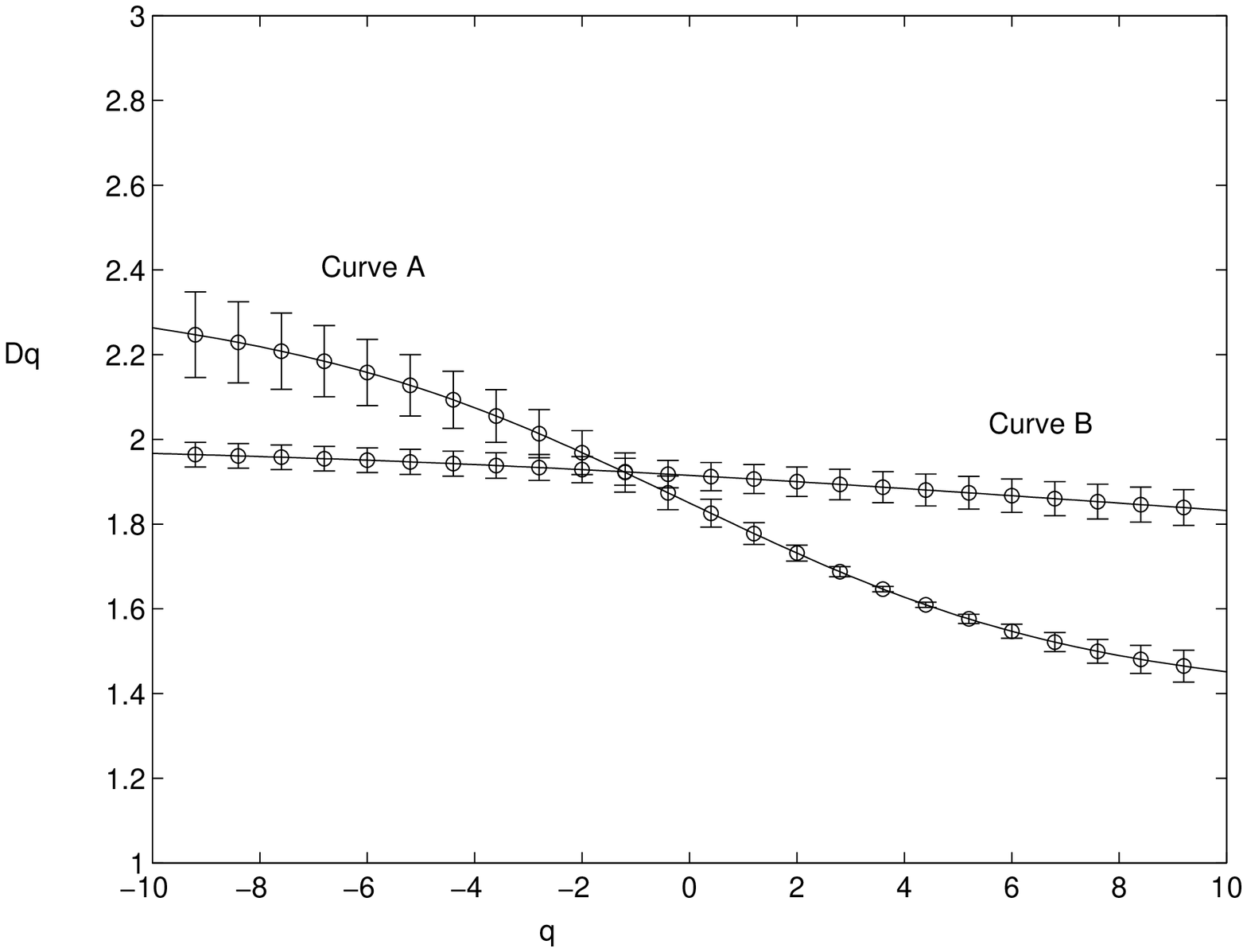}}
\hfill
\caption{The spectrum of generalized dimension is shown for the
subsample d-12.1. Curve A refers to small scales and Curve B to large
scales.}
\label{dq12}
\end{figure}

\begin{figure}[ac2]
\resizebox{7cm}{7cm}{\includegraphics{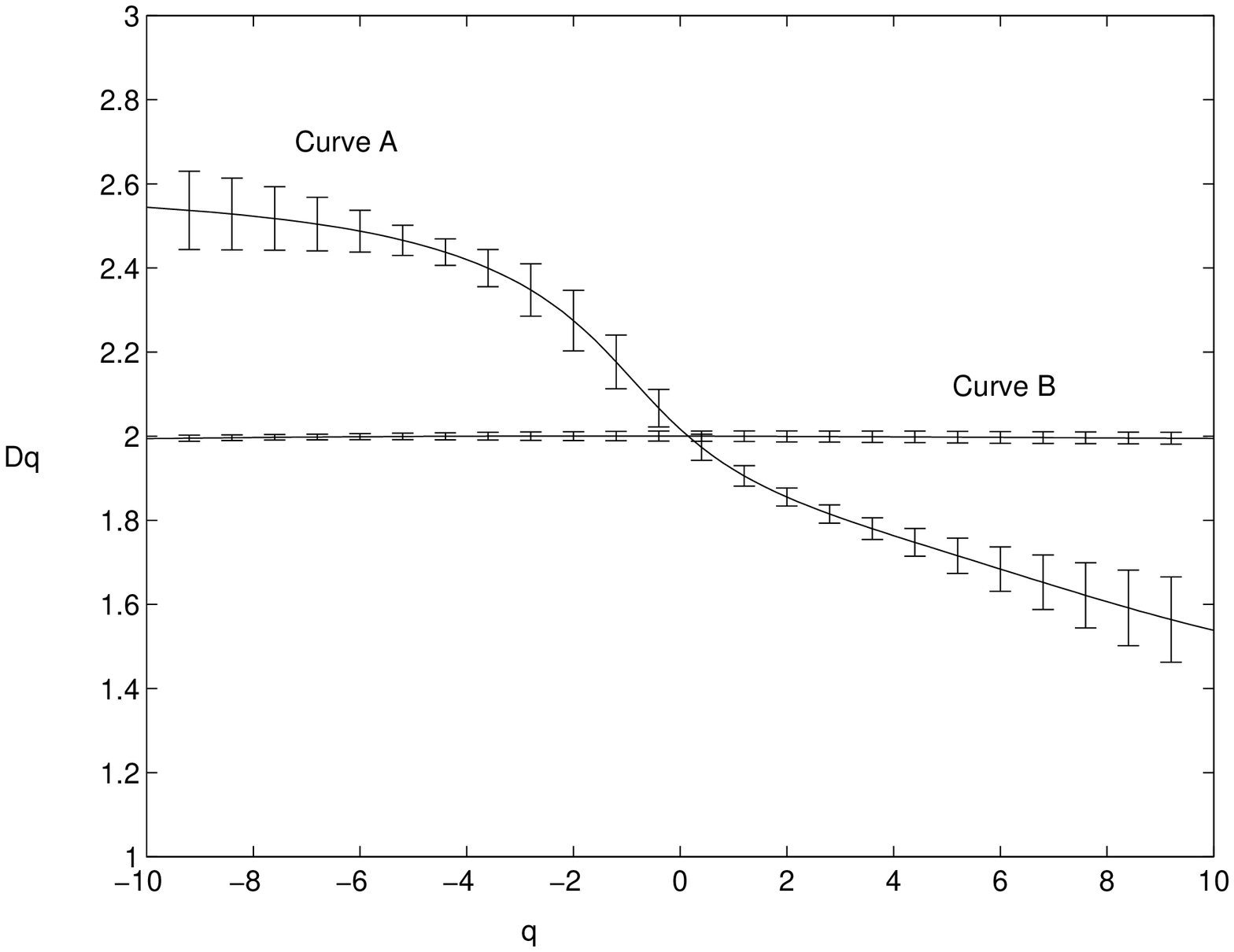}}
\hfill
\caption{The spectrum of generalized dimension is shown for the
subsample d-03.1. Curve A refers to small scales and Curve B to large
scales.}
\label{dq3}
\end{figure}

\begin{figure}[ac3]
\resizebox{7cm}{7cm}{\includegraphics{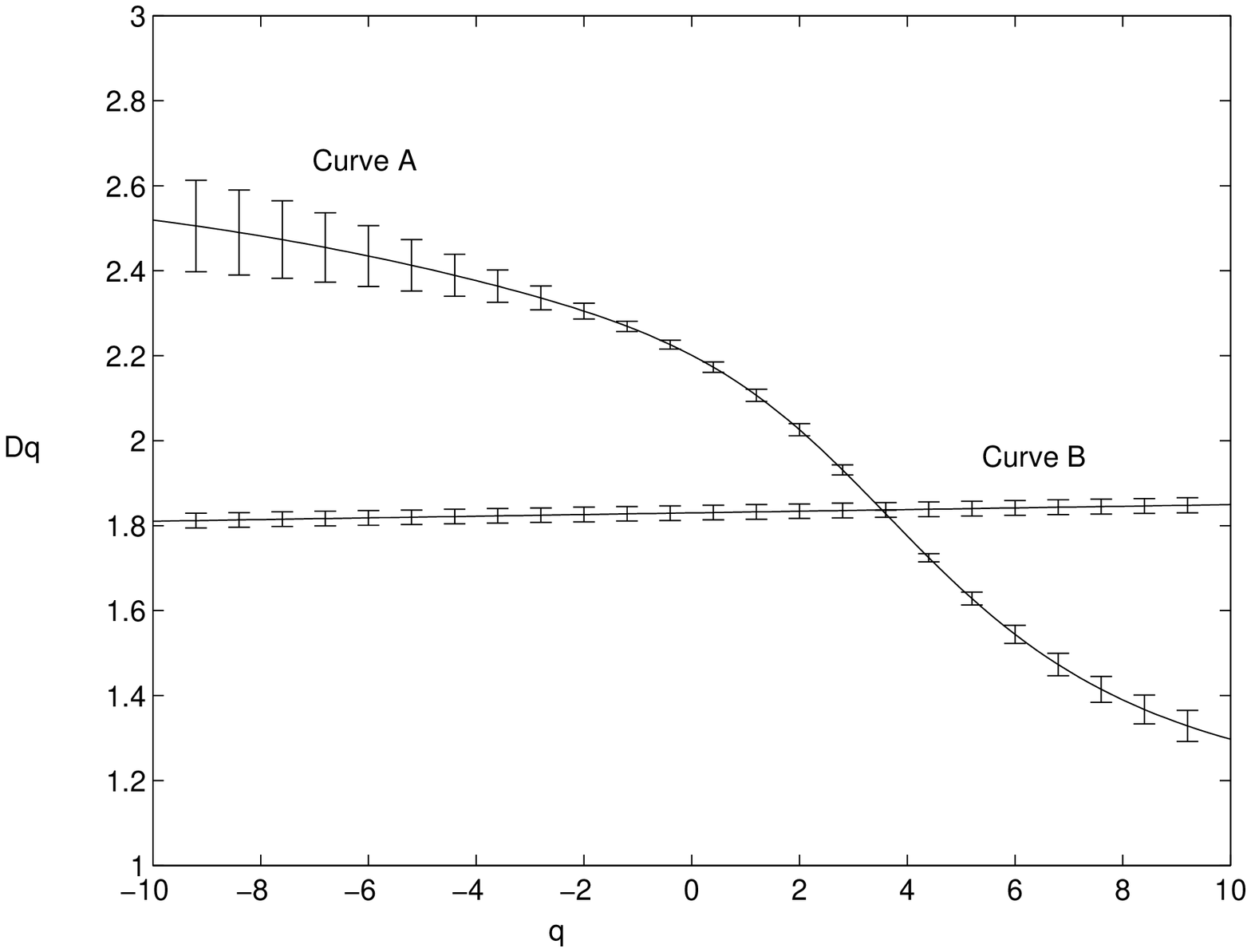}}
\hfill
\caption{The spectrum of generalized dimension is shown for the
subsample d-06.1. Curve A refers to small scales and Curve B to large
scales.}
\label{dq6}
\end{figure}

We next discuss our analysis of the actual data. 
The analysis of the curves corresponding to  $C_q(r)$ versus $r$
for the different subsamples shows the existence of two very
different scaling behaviour - one at small scales and another
at large scales, with the transition occurring  around  $80 h^{-1} {\rm
Mpc}$ to $100 h^{-1} {\rm Mpc}$. The scaling behaviour of $C_q(r)$ is shown in
figure ~\ref{scaleq0} and figure ~\ref{scaleq2} for $q=0$ and $q=2$,
respectively for the
subsample d-12.1 The other subsamples
all exhibit a similar behaviour. Based on this we have treated
the scales $20 h^{-1} {\rm Mpc} \le r \le80 h^{-1} {\rm Mpc}$ (small
scales) and $80 h^{-1} {\rm Mpc} \le r \le 200  h^{-1} {\rm Mpc}$ (large
scales) separately and the  multi-fractal  analysis      
has been  performed separately for the small and large scales.  
Figures ~\ref{dq12}, ~\ref{dq3} and  ~\ref{dq6} show  the spectrum of
generalized  dimensions $D_q \, vs \, q$ at both small and large
scales  for three of the subsamples. 

We find that at small scales  the plots of
$D_q$ versus $q$ for the actual data (figures ~\ref{dq12}, ~\ref{dq3},
~\ref{dq6} ) are quite different from the corresponding plots for the
mock versions of the data (figures ~\ref{mock12}, ~\ref{mock3},
~\ref{mock6}). This clearly shows that the distribution of 
galaxies is not homogeneous over the scales $20h^{-1} \rm{Mpc} \le r
\le 80 h^{-1} \rm{Mpc}$. In addition we find that all the
subsamples exhibit a  multi-fractal behaviour over this range of
length-scales.  The  interpretation of the different values of the
multi-fractal dimension $D_q$ is  complicated by the geometry of the
survey and we do not attempt this here. 

At large scales the behaviour of the generalized dimension $D_q$ is
quite different. For the subsample d-12.1 the spectrum
shows a weak $q$ dependence (figure \ref{dq12}) and $D_q$ shows a
gradual change from 
$D_q \simeq 2$ to $D_q  \simeq 1.8$ as $q$ varies from $-10$ to
$10$. This is quite different from the behaviour at small scales where
the change in $D_q$ is larger and more abrupt. The behaviour of the
other subsamples of the $\delta=-12^{\circ}$ slice are similar. For
the subsample d-03.1 we find that the spectrum is nearly flat 
(figure \ref{dq3}) with $D_q \simeq 2$ and for d-06.1 (figure
\ref{dq6}) the spectrum is nearly flat with $D_q \simeq 1.8$. These
values are within the range we recover from our analysis of the mock
subsamples which are constructed from an underlying random homogeneous
distribution of galaxies. This agreement between the actual data and
the random realizations  with $2.2 \le D_q \le 1.8$ in all the
subsamples  shows  that the distribution of galaxies in LCRS is
homogeneous at the large scales.  

The work presented here contains significant improvements on the
earlier work of Amendola \&  Palladino (1999) on two counts and these
are explained below:\\
(1). Unlike the earlier work which has analyzed volume limited subsamples
of one of the slices ($\delta=-12^{\circ} $) of the LCRS we have
analyzed both volume and magnitude limited subsamples of all the three
northern slices of the  LCRS. The magnitude limited samples contain more
than four times the number of galaxies in the volume limited samples
and they extend to higher redshifts. This allows us to make better use
of the data in the LCRS to improve the statistical significance of the
results and to probe scales  larger than those studied in the previous
analysis. \\
(2). We have calculated the full spectrum of generalized dimensions
which has information about the nature of clustering in different
environments. The integrated conditional density used by the earlier
workers is equivalent to a particular point $(q=2)$ on the spectrum
and it does not fully characterize the scaling properties of the
distribution of galaxies. 

\section{Conclusion.}
\label{conc}
Here we present a method for carrying out  the multi-fractal analysis of
both magnitude and volume limited subsamples of the LCRS.  Our method
takes into account the various selection 
effects and the complicated geometry of the survey. 

We first apply our method to random realizations of the LCRS subsamples for
which we ideally expect a flat spectrum of generalized dimensions with
$D_q =2$. Our analysis gives a nearly flat spectrum with $1.8 \le D_q
\le 2.2$ on large scales. The deviation from the expected value
includes statistical errors arising from the finite number of galaxies
and systematic errors  arising from our treatment of the selection
effects and the complicated geometry. The fact that the errors are
small clearly shows that our method correctly accounts for these
effects. 

Our analysis of the actual data shows the existence of two different
regimes and the  distribution of galaxies on scales $20 h^{-1} {\rm
Mpc}  \le r \le 80 h^{-1} {\rm Mpc}$  shows clear indication of a
multi-fractal  scaling behaviour. On large scales $80 h^{-1} {\rm
Mpc}  \le r \le 200 h^{-1} {\rm Mpc}$ we find a nearly flat spectrum
with $1.8 \le D_q \le 2.2$. This is consistent with our analysis of
the random realizations which have been constructed from a homogeneous
underlying distribution of galaxies. 

Based on the above analysis we conclude that the distribution of
galaxies in the Las Campanas Redshift Survey is homogeneous at large
scales with the transition to homogeneity occurring somewhere around 
$80 h^{-1} {\rm Mpc}$ to $100 h^{-1} {\rm Mpc}$.

\acknowledgements
TRS would like to thank thank T. Padmanabhan, K. Subramanian,
J. S. Bagla, F. S. Labini and   L. Pietronero for several useful
discussions.  AKG and TRS gratefully acknowledge the project grant
(SP/S2/009/94) from the Department of Science and Technology,
India. All the authors are extremely grateful to  the LCRS team for
making  the catalogue publicly available.

\newpage

\newpage

\begin{center}
{\bf Table 1.}
\end{center}

\begin{tabular}{lllllll} \hline
&  & &Absolute&Luminosity&Number of&Vol./Mag.\\ 
subsample&$\,\,\,\delta$&z range& Magnitude range& 
Function&Galaxies &Limited\\\\ \hline 
d-12.1&$-$12.0&0.017-0.2&$-$23.0 - $-$17.5&NS112&4458&M\\
d-12.2&$-$12.0&0.017-0.2&$-$23.0 - $-$17.5&N112&4458&M\\
d-12.3&$-$12.0&0.05-0.1&$-$21.0 - $-$20.0&N112&869 &V\\
d-12.4&$-$12.0&0.065-0.125&$-$21.5 - $-$20.5&N112&923&V\\
d-06.1&$-$6.0&0.017-0.2&$-$23.0 - $-$17.5&NS112&2316&M\\
d-03.1&$-$3.0&0.017-0.2&$-$23.0 - $-$17.5&NS112&4055&M\\
\hline
\end{tabular}
\vspace{.5cm}

\end{document}